\documentclass[12pt]{elsart}
\usepackage[dvips]{color}       
\usepackage{epsfig}

\def\pmb#1{\setbox0=\hbox{#1}
\kern-.025em\copy0\kern-\wd0
\kern.05em\copy0\kern-\wd0
\kern-.025em\raise.0433em\box0}
\def\mbi#1{{\pmb{\mbox{\scriptsize ${#1}$}}}}

\def\bm#1{{\pmb{\mbox{${#1}$}}}}
\def\strut{\vrule width0pt height 15pt depth 7pt}

\begin{document}

\begin{frontmatter}

\title{N$^*$ electroproduction amplitudes in a model with
dynamical confinement}

\author[a,b,EM1]{P. Alberto},
\author[a,b,EM2]{M. Fiolhais},
\author[c,EM3]{B. Golli} and
\author[b,EM4]{J. Marques}

\address[a]{Department of Physics, University of Coimbra,
                 P-3004-516 Coimbra, Portugal}
\address[b]{Centre for Computational Physics, University of
                 Coimbra P-3004-516 Coimbra, Portugal}
\address[c]{Faculty of Education, University of Ljubljana, and
	         J. Stefan Institute, Ljubljana, Slovenia}

\thanks[EM1]{E-mail:pedro@teor.fis.uc.pt} 
\thanks[EM2]{E-mail:tmanuel@teor.fis.uc.pt}
\thanks[EM3]{E-mail:bojan.golli@ijs.si}
\thanks[EM4]{E-mail:jpcmarques@hotmail.com}

\date{\today}

\begin{abstract}
The Roper resonance is described in a chiral version of 
the chromodielectric model as a cluster of three 
quarks in radial-orbital configuration (1s)$^2$(2s)$^1$, 
surrounded by $\pi$ and $\sigma$-meson clouds and by 
a chromodielectric field which assures quark dynamical confinement. 
Radial profiles for all fields are determined self-consistently
for each baryon.
Transverse $A_{1/2}$ and scalar $S_{1/2}$ helicity amplitudes for 
the nucleon-Roper transition are calculated. 
The contribution of glueball and $\sigma$-meson vibrations 
is estimated; although small for N(1440), 
the $\sigma$ contribution can be large for N(1710).\\
\noindent(PACS 12.39.Fe, 13.40.Gp, 14.20.Gk)
\end{abstract}
\end{frontmatter}

The new facilities for intermediate energy nuclear physics, 
operating with continuous electron beams, make more accessible
accurate measurements of electromagnetic properties of both 
the nucleon and excited states, thus providing more and better 
information on the structure  of baryons, and stimulating  
theoretical research on the structure of the nucleons and 
its resonances. 
The Roper resonance, N(1440), is of particular interest since, 
due to its relatively low excitation energy,
a simple picture in which one quark populates 
the 2s level does not work here. 
The constituent quark model (CQM) does not yield sensible
results for the electromagnetic properties unless the quark
dynamics is treated relativistically~\cite{Capstick,Weber}
and, furthermore,
approximations beyond the simple Gaussian approximation~\cite{Cardarelli},
or inclusion of $q\bar{q}$ pairs \cite{Cano} are taken into account.
These difficulties suggest that additional degrees of freedom,
such as explicit excitations of glue-field~\cite{Li},
glueball field~\cite{broRPA}, or chiral fields~\cite{Krehl,dong,tiator}
may be important for formation of the Roper resonance.

In this letter we use a simple model, the chromodielectric
model (CDM),  which is particularly suitable to 
describe the interplay of glueball and meson excitations together 
with the usual quark radial excitation.
In contrast to the nonrelativistic or relativistic versions
of the constituent quark model, in the CDM the electromagnetic 
current operator is derived directly from the Lagrangian, hence 
no additional assumptions have to be introduced in 
the calculation of electroexcitation amplitudes.
The electromagnetic current contains an explicit contribution from 
the pion field which has been shown to play an important role
in the description of the N--$\Delta$ electroproduction
\cite{delta1}.

The Roper has been considered in a non-chiral version of 
the CDM using the RPA techniques to describe coupled vibrations 
of valence quarks and the background chromodielectric 
field~\cite{broRPA}.
The energy of the lowest excitation turned out to be
40~\% lower than the pure 1s--2s excitations.
A similar result was obtained by Guichon~\cite{guiMIT}, 
using the MIT bag model and considering the Roper as a 
collective vibration of valence quarks and the bag.

In our approach we describe the nucleon and the Roper
as chiral solitons resulting from the non-linear 
interactions between quarks and a scalar-isoscalar chiral
singlet field $\chi$ which, through the peculiar way it couples 
to the quarks, provides a mechanism for confinement.
In addition, the quarks interact with scalar-isoscalar ($\sigma$)
and pseudoscalar-isovector ($\pol{\pi}$) mesons
similarly as in the linear $\sigma$-model, though
in the CDM the chiral fields are  weaker and similar to the 
solution in the CBM for bag radius above 1~fm.
The Lagrangian of the model can be written as \cite{BirseP}
\begin{equation}
  \mathcal{L} = \mathcal{L}_q + \mathcal{L}_{\sigma,\pi} 
    +  \mathcal{L}_{q-\mathrm{meson}} +  \mathcal{L}_\chi\;,
  \label{langrangian}
\end{equation}
where
\begin{equation}
  \mathcal{L}_q = \mathrm{i}\bar{\psi}\gamma^\mu \partial_\mu\psi \;,
\qquad
  \mathcal{L}_{\sigma,\pi} =
  \half\partial_\mu\hat{\sigma}\partial^\mu\hat{\sigma}
  + \half\partial_\mu\hat{\pol{\pi}}\cdot\partial^\mu\hat{\pol{\pi}} 
  - \mathcal{U}(\hat{\pol{\pi}}^2+\hat{\sigma}^2)\;,
  \label{langrangian1}
\end{equation}
$\mathcal{U}(\hat{\pol{\pi}}^2+\hat{\sigma}^2)$ being the usual
Mexican hat potential, and the quark meson interaction is given by
\begin{equation}
    \mathcal{L}_{q-{\mathrm{meson}}} = {g\over\chi}\, \bar{\psi}
    (\hat{\sigma}+\mathrm{i}\pol{\tau}\cdot\hat{\pol{\pi}}\gamma_5)
    \psi\;.
   \label{langrangian2}
\end{equation}
The last term in (\ref{langrangian})
contains the kinetic and the potential piece for the $\chi$-field:
\begin{equation}
  \mathcal{L}_\chi =  
  \half\partial_\mu\hat{\chi}\,\partial^\mu\hat{\chi}
  - {1\over2}M^2\,\hat{\chi}^2\;.
  \label{langrangian3}
\end{equation}
Other versions of the CDM consider a quartic potential 
in (\ref{langrangian3}).
By taking just the mass term the confinement is imposed
in the smoothest way, which seems to be the most appropriate 
choice for the quark matter sector of the CDM \cite{Dra95}.

The parameters of the model have been fixed by requiring
that the calculated static properties of the nucleon agree best 
with the experimental values:
we take $g=0.03$~GeV and $M=1.4$~GeV \cite{delta1,Dra95,drago2}. 
The pion decay constant and the chiral meson masses   
are fixed to $f_\pi=0.093$~GeV and $m_\pi=0.14$~GeV, while for
the mass of the $\sigma$-meson we consider 
values between $m_\sigma=0.7$~GeV and $m_\sigma=1.2$~GeV.
We have checked that our results depend very 
weakly on the variations of these parameters.

The starting point to describe a baryon is the hedgehog coherent state, 
which we write in the form:
\begin{eqnarray}
|Hh  \rangle &=& 
   N \, {\rm exp} \left\{ \sum_{tm} (-1)^{m} \delta_{t,-m}
  \int_0^\infty \d k \sqrt{2 \pi \omega_\pi(k) \over 3} \xi(k) 
  a^\dagger_{tm} (k) \right\}\times     
\nonumber \\ &&    
  {\rm exp} \left\{ \int_0^\infty \d k \sqrt{2 \pi \omega_\sigma(k) } 
   \eta(k) \tilde{a}^\dagger (k)\right\}\times     
\nonumber \\ &&
   {\rm exp}\left\{ \int_0^\infty \d k \sqrt{2 \pi \omega_\chi(k) } 
   \zeta(k) {b}^\dagger (k)\right\}
   \times \prod_{i=1,3} c^\dagger_h (i) |0\rangle \, .
\label{eq:3}
\end{eqnarray}
Here $a^\dagger_{tm}(k)$ is the creation operator for a p-wave pion with 
isospin and angular momentum third components $t$ and $m$ respectively, 
orbital wave function $\xi(k)$ and frequency 
$\omega_\pi=\sqrt{k^2+m_\pi^2}$; 
similarly, $\tilde{a}^\dagger (k)$ and ${b}^\dagger (k)$ create 
s-wave  $\sigma$ and $\chi$ quanta, with orbital wave functions 
$\eta(k)$ and $\zeta(k)$, 
and frequencies $\omega_\sigma$ and $\omega_\chi$, respectively; 
$N$ is a normalization constant.
The amplitudes, $\xi(k)$, $\eta(k)$ and $\zeta(k)$, are 
Fourier transforms of the corresponding pion, $\sigma$ and $\chi$ radial 
profiles, which we denote as $\phi(r)$, $\sigma(r)$, and $\chi(r)$,
respectively. 
Finally, the operator $c_h^\dagger(i)$ creates a s-wave valence 
quark in a spin-isospin hedgehog state:
\begin{equation}
\!\!\!\! \langle{\bm r} | c^\dagger_h(i) |0\rangle 
   = q_i({\bm r})
   = {1 \over \sqrt{4 \pi}} \left( 
   \begin{array}{c}
      u_i(r)  \\ {\rm i} v_i(r) \, {\bm \sigma} \cdot \hat{\bm r}
    \end{array} \right)|h\rangle \, , 
\quad  
   |h\rangle\!=\!{1\over\sqrt{2}} 
    \left(|u\!\!\downarrow\rangle - |d\!\! \uparrow\rangle \right) \, . 
\label{eq:::10}
\end{equation}
The index $i$ distinguishes between different radial states.  
The physical states are obtained by performing 
the Peierls-Yoccoz projection~\cite{GR85}:
\begin{equation}
|{\rm N}_{{1\over 2},M_T}\rangle 
   = \mathcal{N}\,
     P^{1\over 2}_{{1\over 2},-M_T} |Hh\rangle \, , 
\quad 
    |{\rm R}'_{{1\over 2},M_T}\rangle 
     = \mathcal{N}'\,
     P^{1\over 2}_{{1\over 2},-M_T} |Hh^*\rangle \, .   
\label{usa}
\end{equation}
The $\chi$ and the $\sigma$-fields are not affected by projection.
The radial profiles $\phi(r)$, $\sigma(r)$ and $\chi(r)$, and the  
quark profiles are determined self-consistently
using variation after projection~\cite{BirseP}, separately for the
nucleon and for the Roper.
For the Roper, one has still to distinguish between the radial 
functions for quarks in 1s state and in 2s state, 
and we shall use the self-explanatory notation $u^*_1$, $v^*_1$, 
$u^*_2$, $v^*_2$, $\sigma^*$, $\phi^*$ and $\chi^*$.  
The states (\ref{usa}) are normalized but not mutually
orthogonal. They can be orthogonalized by taking
\begin{equation}
   |{\rm R}\rangle 
    = {1\over\sqrt{1-c^2}}(|{\rm R}'\rangle - c|{\rm N}\rangle)\;,
\qquad
  c = \langle{\rm N}|{\rm R}'\rangle \;.
\label{orthoR}
\end{equation}
A better procedure
results from a diagonalization of the Hamiltonian in the subspace 
spanned by $|{\rm R}'\rangle$ and $|{\rm N}\rangle$:
\begin{equation}
 |\tilde{\rm R}\rangle = c^R_R|{\rm R}'\rangle + c^R_N|{\rm N}\rangle\,, 
\quad
 |\tilde{\rm N}\rangle = c^N_R|{\rm R}'\rangle + c^N_N|{\rm N}\rangle\,. 
\label{gcmss}
\end{equation}

A central point in our treatment of the Roper 
is the freedom of the chromodielectric profile, 
as well as of the chiral meson profiles, 
to adapt to a  (1s)$^2$(2s)$^1$ configuration. 
Therefore, quarks in the Roper experience mean fields 
which are different from the mean boson fields felt by the quarks 
in the nucleon. 

\begin{figure}[b]
\centerline{\epsfig{clip=on,file=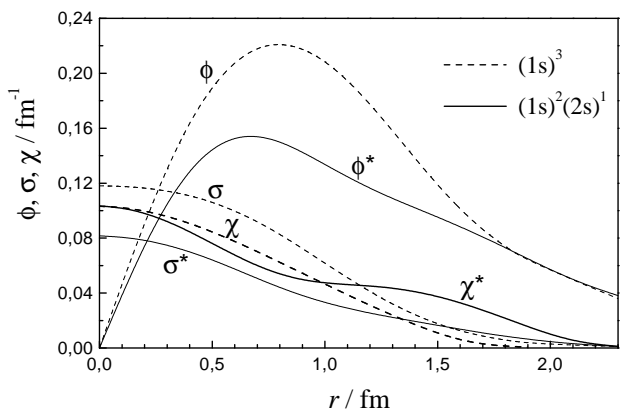,width=9cm} }
\caption{Self-consistent chromodielectric field  and 
chiral (pion and sigma) fields for quark configurations (1s)$^3$ ($\chi$) 
and (1s)$^2$(2s)$^1$ ($\chi^*$). The represented sigma profile is its 
fluctuation from the vacuum $-f_\pi$.
We used the parameter set $g=0.03$~GeV, $M=1.4$~GeV 
and $m_\sigma=0.85$~GeV.}
\label{fig:fig1}
\end{figure}

\begin{figure}[bht]
\centerline{\epsfig{clip=on,file=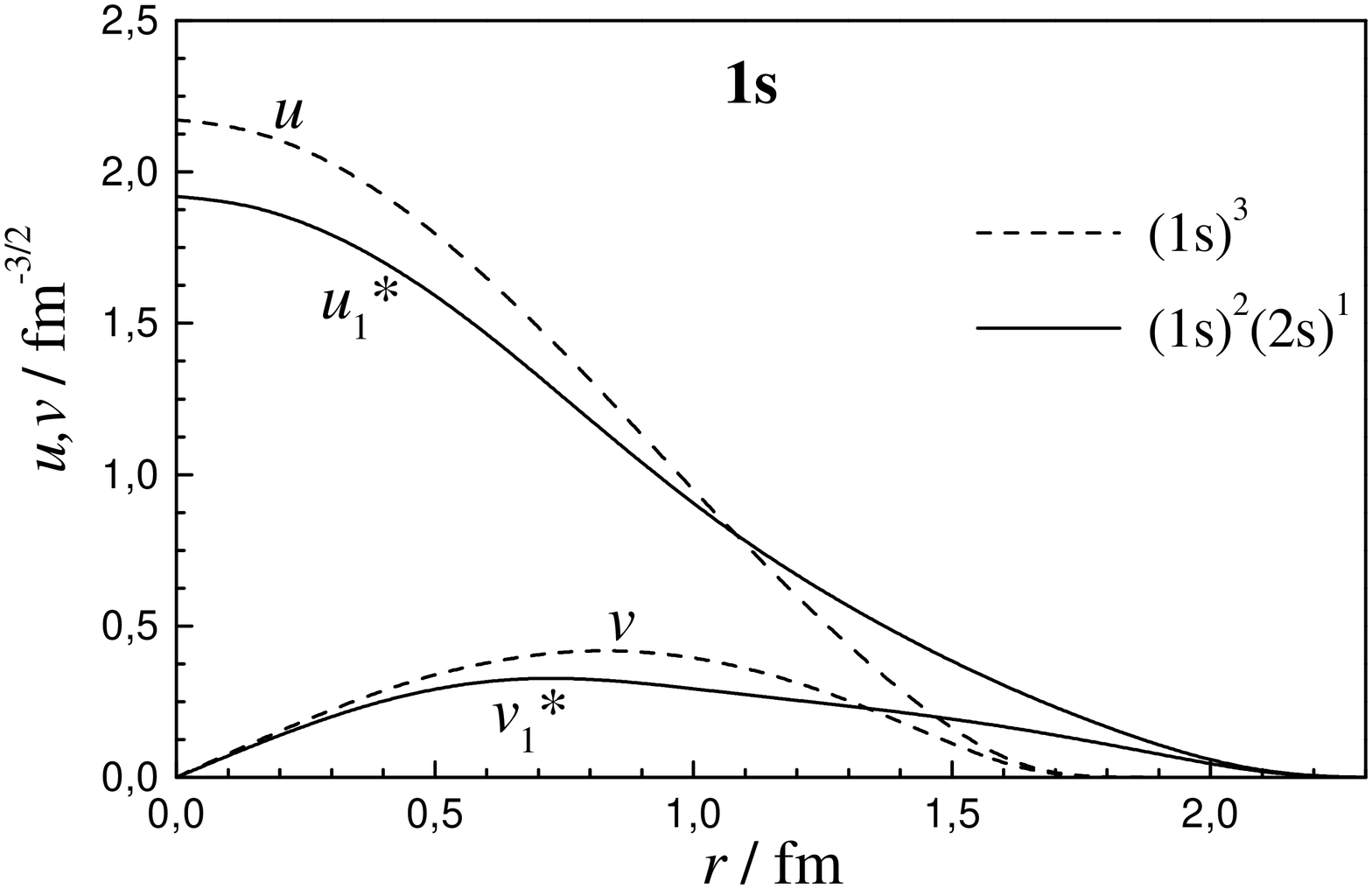,width=9cm}}
\centerline{\epsfig{clip=on,file=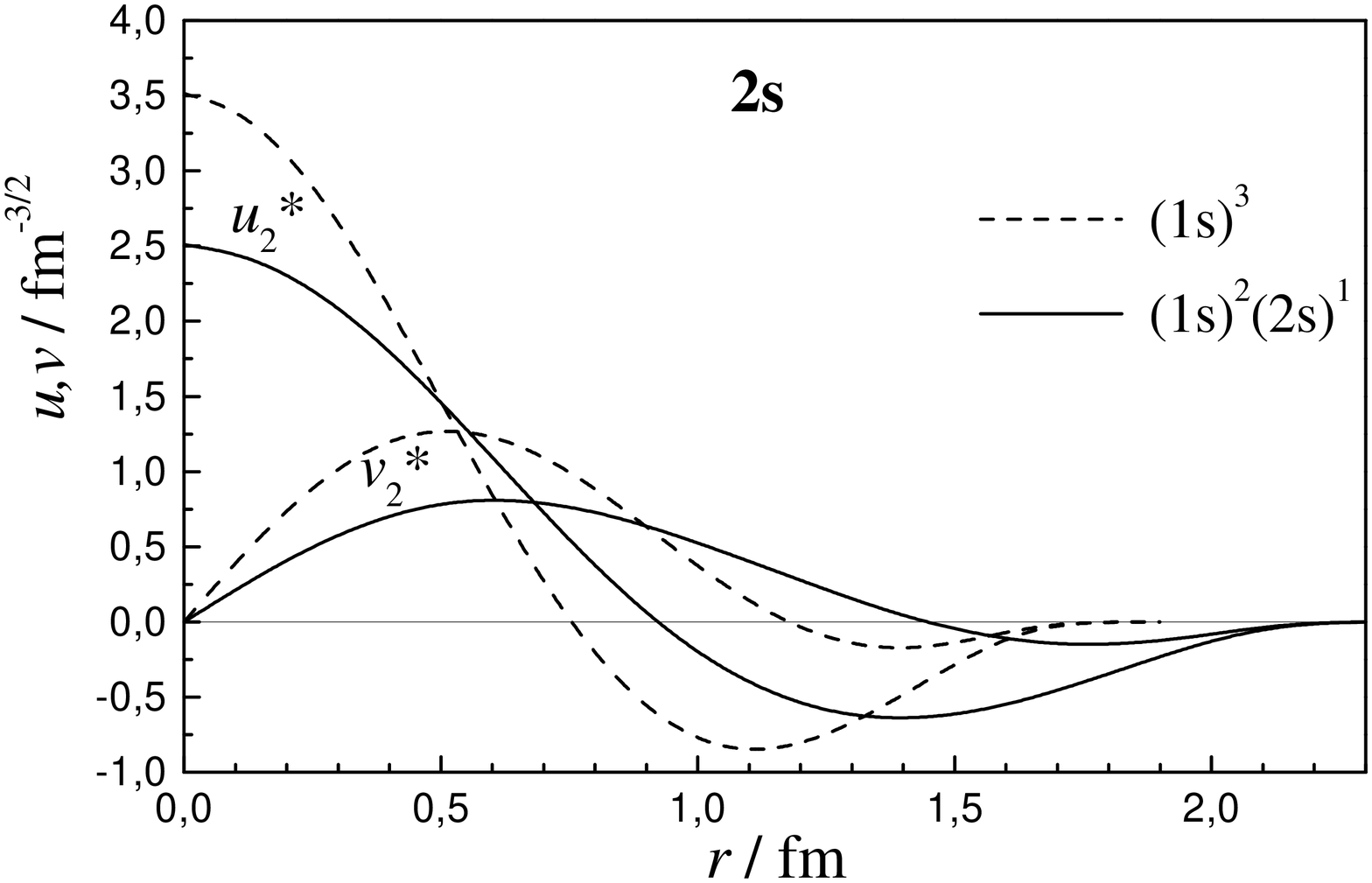,width=9cm}}
\caption{Quark radial profiles of 1s state and 2s states for (1s)$^3$ 
and (1s)$^2$(2s)$^1$ configurations. The dashed curves on the lower  panel 
were computed using frozen ground-state meson fields. 
The full curves in the lower panel as well
as all curves in the upper panel were determined self-consistently. 
Same parameters as in Fig.~\ref{fig:fig1}.}
\label{fig:fig2}
\end{figure}

The self-consistently determined fields shown in Fig.~\ref{fig:fig1} 
depend noticeably on the quark source configuration.  
The $\chi$-field is almost insensitive to the $\sigma$-meson mass
while the strength of the chiral fields increase with 
decreasing $\sigma$-meson mass.
In Fig.~\ref{fig:fig2} we show the quark radial profiles 
calculated self-consistently and, for the 2s state, also the 
radial profile calculated with the fixed background fields 
determined in the self-consistent calculation of the nucleon 
ground state (hereafter we call this approximation
``frozen fields" calculation and denote it by ``ff" to distinguish it 
from the self-consistent calculation, denoted by ``sc").
In Table~\ref{tab:1} we give the nucleon energy, $E_N$, 
and the Roper--nucleon energy splitting for various approximations.
The 1s--2s level splitting corresponds to the ff calculation 
(in this case the boson fields do not contribute to the difference),  
while $\Delta E_R$ refers to the self-consistent calculations.  
In the latter case, the level splitting itself is reduced by 35~\%
with respect to ff,
but the total Roper--nucleon splitting is actually higher due to 
the increase of the potential energy of the fields and to 
the orthogonalization (\ref{orthoR}).

The ansatz (\ref{usa}) for the Roper represents the breathing mode 
of the three valence quarks with the fields adapting to the
change of the source.
There is another possible type of excitation in which
the quarks remain in the ground state
while the $\chi$-field and/or the $\sigma$-field  oscillate.
Such oscillations can be simply described if the ground state 
is taken in the form of a (projected) coherent state.
We expand the field operators of the scalar bosons 
around  their expectation value in the ground state
$|{\rm N}\rangle$ (\ref{usa}):
\begin{equation}
   \hat{\sigma}(\vec{r}) =
       \sum_n{1\over\sqrt{2\varepsilon_n}}\, \varphi_n(r)
        {1\over\sqrt{4\pi}}
        \left[\tilde{a}_n + \tilde{a}^\dagger_n\right] + \sigma(r)
\;,
\label{fieldop}
\end{equation}
where the operators $\tilde{a}_n$ annihilate the ground state, 
{\em i.e.} $\tilde{a}_n|N\rangle = 0$.
(We write here explicitly only the vibrations of the $\sigma$-meson;
for the $\chi$ field the derivation is analogous.) From
$\tilde{a}(k)|N\rangle  = \sqrt{2\pi\omega_\sigma(k)}\eta(k)|N\rangle$
one can obtain a simple expression for the anni\-hi\-la\-tion 
(creation) operator of the $n$-th mode\footnote{%
The most general expression would involve the Bogoljubov
transformation; however, the corresponding ground state
would not be a simple coherent state.
If we want to keep the simple ansatz which allows us
to perform calculations of matrix elements, the expression
is the most general transformation that preserves the ground state.
Performing the Bogoljubov transformation leads to RPA
equations; the present approach is therefore a simplified
treatment of RPA excitations.}: 
\begin{equation}
 \tilde{a}_n = \int\d k\, \tilde{\varphi}_n(k)
    \left(\tilde{a}(k) - \sqrt{2\pi\omega_\sigma(k)}\,\eta(k)\right)
\;,
\end{equation}
where $\tilde{\varphi}_n(k)$ is the Fourier transform of the $n$-th 
mode in (\ref{fieldop}).

The stability conditions for  the ground state require that the
$\varphi_n$ and $\varepsilon_n$
satisfy the Klein-Gordon equation:
\begin{equation}
 \left(-\nabla^2 + m^2 
     + {\d^2 V(\sigma(r))\over\d\sigma(r)^2}\right)\varphi_n(r)
  = \varepsilon_n^2 \varphi_n(r)\;.
\label{Klein-Gordon}
\end{equation}
Here $V$ stands for the potential originating from the term
$\mathcal{L}_{q-{\mathrm{meson}}}$ and the potential parts of
$\mathcal{L}_\chi$ and $\mathcal{L}_{\sigma,\pi}$ 
in (\ref{langrangian}). 
For the self-consistently determined profiles of the 
ground state we find that the potential in (\ref{Klein-Gordon})
is {\em repulsive\/} for the $\chi$-field and {\em attractive\/} 
for the $\sigma$-field.
This means that there are no glueball excitation in which the 
quarks would act as spectators: the $\chi$- field oscillates
only together with the quark field.
On the other hand, the effective $\sigma$-meson potential supports 
at least one bound state with the energy $\varepsilon_1$ of 
typically 100~MeV below the $\sigma$-meson mass\footnote{%
The appearance of an attractive potential is not only a
feature of the CDM, it appears in other chiral models in
which the chiral fields are not constrained to the chiral circle;
{\em e.g.} in the recent calculation of the nucleon in the Nambu 
Jona-Lasinio model with nonlocal regulators~\cite{GBR} the chiral 
fields in the center of the soliton turn out to be quite far away 
from the chiral circle, producing a strong attractive potential
for the $\sigma$-meson.
} (see Table~\ref{tab:1}).
The lowest excited state is obtained by populating
the lowest mode of the vibrator with one boson.

We can now  extend the ansatz  (\ref{orthoR}) by introducing
\begin{equation}
  |{\rm R}^*\rangle = c_1|{\rm R}\rangle 
       + c_2\tilde{a}^\dagger_\sigma|N\rangle\;,
\label{gRstate}
\end{equation} 
where $\tilde{a}^\dagger_\sigma$ is the creation operator 
for this lowest vibrational mode.
The coefficients $c_i$ and the energy are determined 
by solving the (generalized) eigenvalue problems 
in the $2\times2$ subspace.
The solution with the lowest energy corresponds to the Roper
while the orthogonal combinations could be attributed
to the second excited state with nucleon quantum numbers,
the $N(1710)$, provided the $\sigma$-meson mass
is sufficiently small.
In such a case the latter state is described as
predominantly the $\sigma$-meson vibrational mode
rather than the second radial excitation of quarks.

The energy of the Roper resonance is reduced when the lowest 
vibrational mode is included in the ansatz; the reduction is 
small due to the small coupling between the state (\ref{orthoR})
and the lowest vibrational state with the energy $\varepsilon_1$
(see Table~\ref{tab:1}).                   
The orthogonal combination is practically at 
energy $E_N+\varepsilon_1$.
The orthogonalization procedure (\ref{gcmss}) lowers the 
ground-state energy and increases the nucleon--Roper energy 
splitting with respect to $\Delta E_R$.

\begin{table}[h]
\begin{center}
\begin{tabular}{rrrrrrrrr}
\hline

\hline
\strut $m_\sigma$ & $\ \ \ \ E_N$ & 2s--1s & $\Delta E_R$ & 
$\ \ \ \ \varepsilon_1$ & $\Delta E_{R*}$ & $c_2$ & 
   \ \ \ \  $\tilde{E}_N$ & $\Delta \tilde{E}_R$\\
\hline
\strut 1200  & 1269  & 446 &  354 &  1090 & 353 & 0.05 & 1256 &  380 \\
\strut   700 &  1249  & 477 &  367 &   590 & 364 & 0.12 & 1235 &  396 \\
\hline

\hline 
\end{tabular}
\vskip0.2cm
\caption{For two sigma masses, listed are the nucleon energy using
(\ref{usa}) ($E_N$) and (\ref{gcmss}) ($\tilde{E}_N$),
the nucleon-Roper energy difference for fixed background fields (2s--1s), 
for the state (\ref{orthoR}) ($\Delta E_R$),
for  (\ref{gRstate}) ($\Delta E_{R*}$),
and for (\ref{gcmss}) ($\Delta \tilde{E}_R$); 
$\varepsilon_1$ is the energy of the lowest vibrational mode 
and $c_2$ its strength.
All energies are in MeV.}
\label{tab:1}
\end{center}
\end{table}

We now turn to the presentation of the electromagnetic 
nucleon--Roper transition amplitudes.
Using the state vectors (\ref{gcmss}) the nucleon--Roper resonant 
electromagnetic transition amplitudes are readily evaluated.
One usually introduces the resonant transverse helicity amplitude
defined in the rest frame of the resonance as 
\begin{equation}
A_{1/2}=- \zeta\,\sqrt{2 \pi \alpha \over k_W}  \int \d^3 {\bm r} \, \
  \langle\tilde{\rm R}_{+{1\over 2}, M_T}|{\bm J}_{\rm em}({\bm r})\cdot 
   {\bm \epsilon}_{+1} \, {\rm e} ^{{\rm i}{\mbi k} 
     \cdot {\mbi r} } |\tilde{\rm N}_{-{1\over 2}, M_T }\rangle
\label{helicea}
\end{equation}
and a scalar helicity amplitude
\begin{equation}
S_{1/2}=\zeta\,\sqrt{2 \pi \alpha \over k_W}  \int \d {\bm r}  \, 
    \langle\tilde{\rm R}_{+{1\over 2}, M_T}|  J^0_{\rm em} ({\bm r}) \,\, 
     {\rm e} ^{{\rm i}{\mbi k} 
     \cdot {\mbi r} } |\tilde{\rm N}_{+{1\over 2}, M_T}\rangle\, .
\label{helices}
\end{equation}
where $\alpha={e^2\over 4\pi}={1\over 137}$ is 
the fine-structure constant, 
the unit vector ${\bm \epsilon}_{+1}$ is the polarization vector of 
the electromagnetic field,  $k_W$ is the photon momentum 
at the photon point, 
$
  k_W=(M_{\rm R}^2 - M_{\rm N}^2) / 2 M_{\rm R}
$,
and $\zeta$ is the sign of the $N\pi$ decay amplitude~\cite{Babcock}:
$\zeta = {\rm sgn}\,\left\{{\langle R(\half\half) 
\rightarrow
\pi_a + N_b(M=-\half)\rangle/(1a\,\half b|\half\half)}\right\}$,
where $a$ and $b$ are the third components of pion and
nucleon isospin, respectively.
This sign 
has to be explicitly calculated within the model.
Performing the multipole decomposition of the (transverse) 
electromagnetic field, 
${\bm A}_{+1}\!=\! {\bm \epsilon}_{+1} 
{\rm e}^{{\rm i}{\mbi k} \cdot {\mbi r}}$,
and using the Wigner-Eckart theorem, Eq.~(\ref{helicea}) becomes 
\begin{equation}
A_{1/2} =  \zeta\,\sqrt{{\pi \alpha \over k_W}}
\int \d^3 {\bm r}  {3 j_1(kr) \over r}   
 \langle\tilde{\rm R}_{{1\over 2}, M_T}|\left[ {\bm r} 
  \times {\bm J}_{\rm em}({\bm r})\right]_{0}
  | \tilde{\rm N}_{{1\over 2}, M_T} \rangle\;.  
\label{hela}
\end{equation}
Similarly, Eq.~(\ref{helices}) becomes 
\begin{equation}
  S_{1/2} =\zeta\, \sqrt{{2 \pi \alpha \over k_W}}
  \int \d^3 {\bm r}  j_0(kr)    
   \langle\tilde{\rm R}_{{1\over 2}, M_T}| J^0_{\rm em}({\bm r})
   |\tilde{\rm N}_{{1\over 2}, M_T} \rangle\;.
\label{hels}
\end{equation}

The electromagnetic operator in the CDM, 
$J^\mu_{\rm em}\equiv (J^0_{\rm em},{\bm J}_{\rm em})$, 
contains a quark part and a pion part and reads:
\begin{equation}
  J^\mu_{\rm em} = 
    \sum_{i=1}^3 \overline{q}_i \gamma^{\mu,(i)} \left( {1 \over 6} 
  + {1 \over 2} \tau_0^{(i)} \right) q_i +
   \left(\vec \pi \times \partial ^\mu \vec \pi \right)_0 
\label{emcur}
\end{equation}
where the index 0 refers to the isospin third component.

\begin{figure}[b]
\centerline{\epsfig{clip=on,file=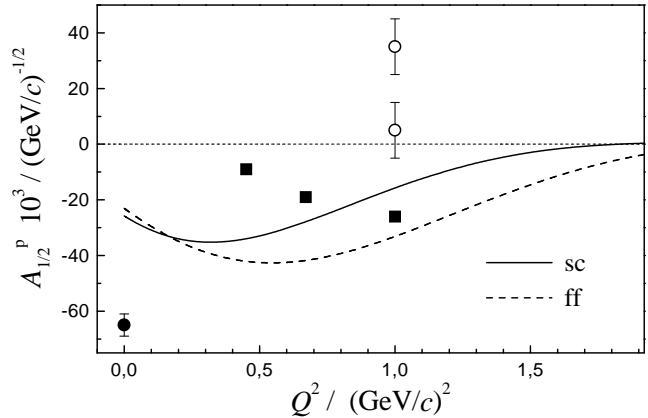,width=9.5cm} }
\caption{Nucleon Roper transverse helicity amplitude for charged 
states. 
The experimental point at $Q^2=0$ is the estimate of the PDG~\cite{PDG}. 
The solid squares result from the analysis of electroproduction 
data performed in \cite{analy1}. 
The open circles also result from an analysis of electroproduction 
data~\cite{analy2}. 
The curves refer to self-consistent (sc) and frozen fields (ff) calculations. 
Parameter set as in Fig.~\ref{fig:fig1}.}\label{fig:fig3}
\end{figure}

\begin{figure}[thb]
\centerline{\epsfig{clip=on,file=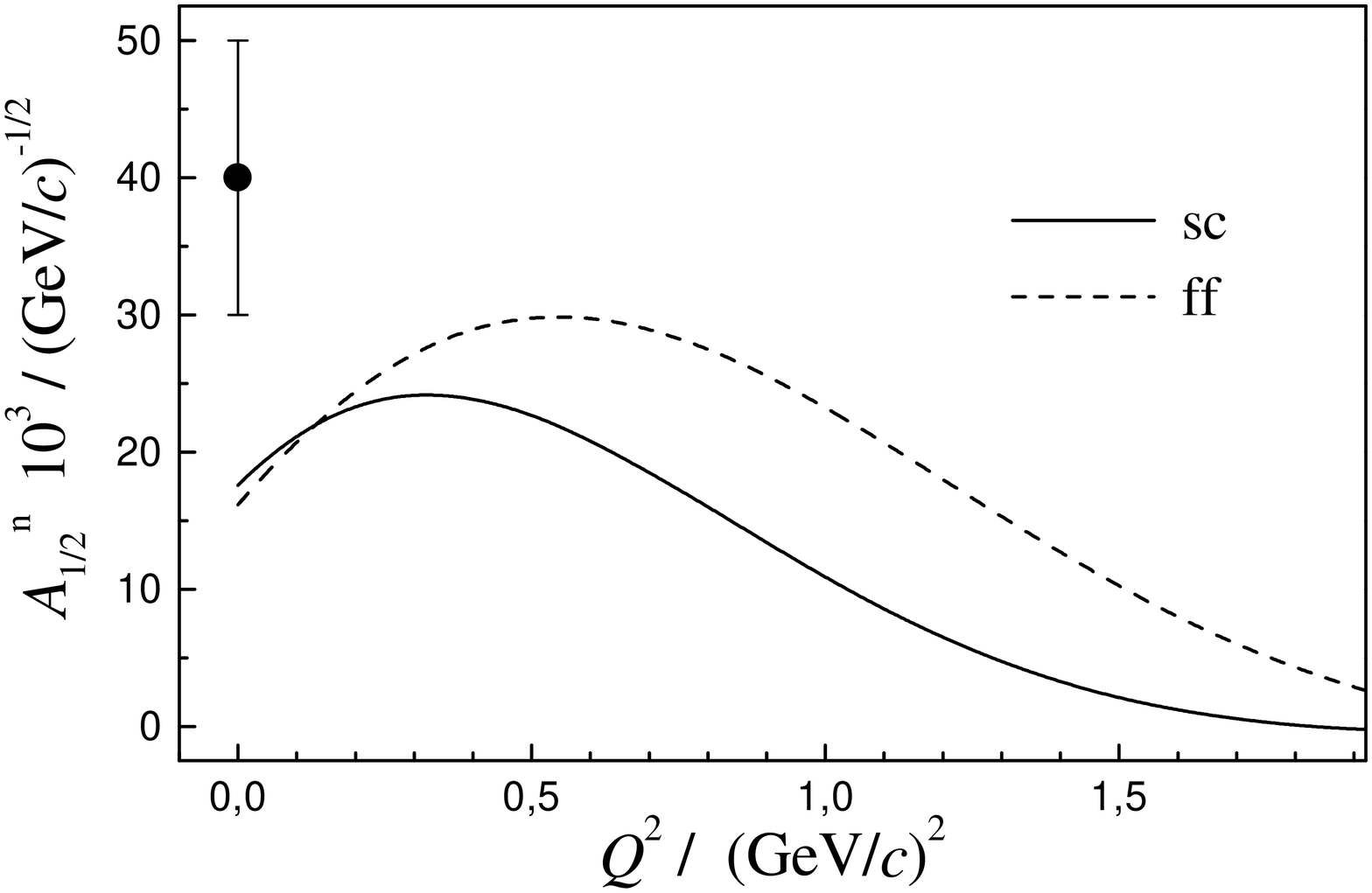,width=9.5cm} }
\caption{Nucleon Roper transverse helicity amplitude for  neutral states.
See also caption of Fig.~\ref{fig:fig3}.}\label{fig:fig4}
\end{figure}

Our results for the transverse helicity amplitudes for the charged 
and neutral states are shown in figs.~\ref{fig:fig3} and \ref{fig:fig4} for 
the parameter set $g=0.03$~GeV, $M=1.4$~GeV and $m_\sigma=0.85$~GeV 
(the amplitudes for $m_\sigma=0.7$~GeV and $m_\sigma=1.2$~GeV
are quantitatively similar to those shown 
in figs.~\ref{fig:fig3} and \ref{fig:fig4}).
Shown are the model predictions for the self-consistent calculations 
(using states (\ref{gcmss})) and for ground state frozen fields 
(using states (\ref{usa})). 
The experimental values at the photon point 
are the PDG most recent estimate \cite{PDG} 
$A_{1/2}^p=-0.065\pm 0.004$~(GeV/$c$)$^{-1/2}$ and  
$-0.040\pm 0.010$~(GeV/$c$)$^{-1/2}$ for $A_{1/2}^n$. 
The pion contribution to the charged states only accounts for 
a few percent of the total amplitude. 
The large discrepancy at the photon point can be
attributed to a too weak pion field in the model
which we already noticed in the calculation 
of nucleon magnetic moments~\cite{drago2} and
of the electroproduction of the $\Delta$~\cite{delta1}.
Other chiral models~\cite{tiator}
predict a much stronger pion contribution which enhances
the value of the amplitudes at the photon point.
If we calculate perturbatively the leading pion contribution
we also find a strong enhancement at the photon point; however,
when we properly orthogonalize the state with respect to the nucleon,
this contribution almost disappears.

\begin{figure}[thb]
\centerline{\epsfig{clip=on,file=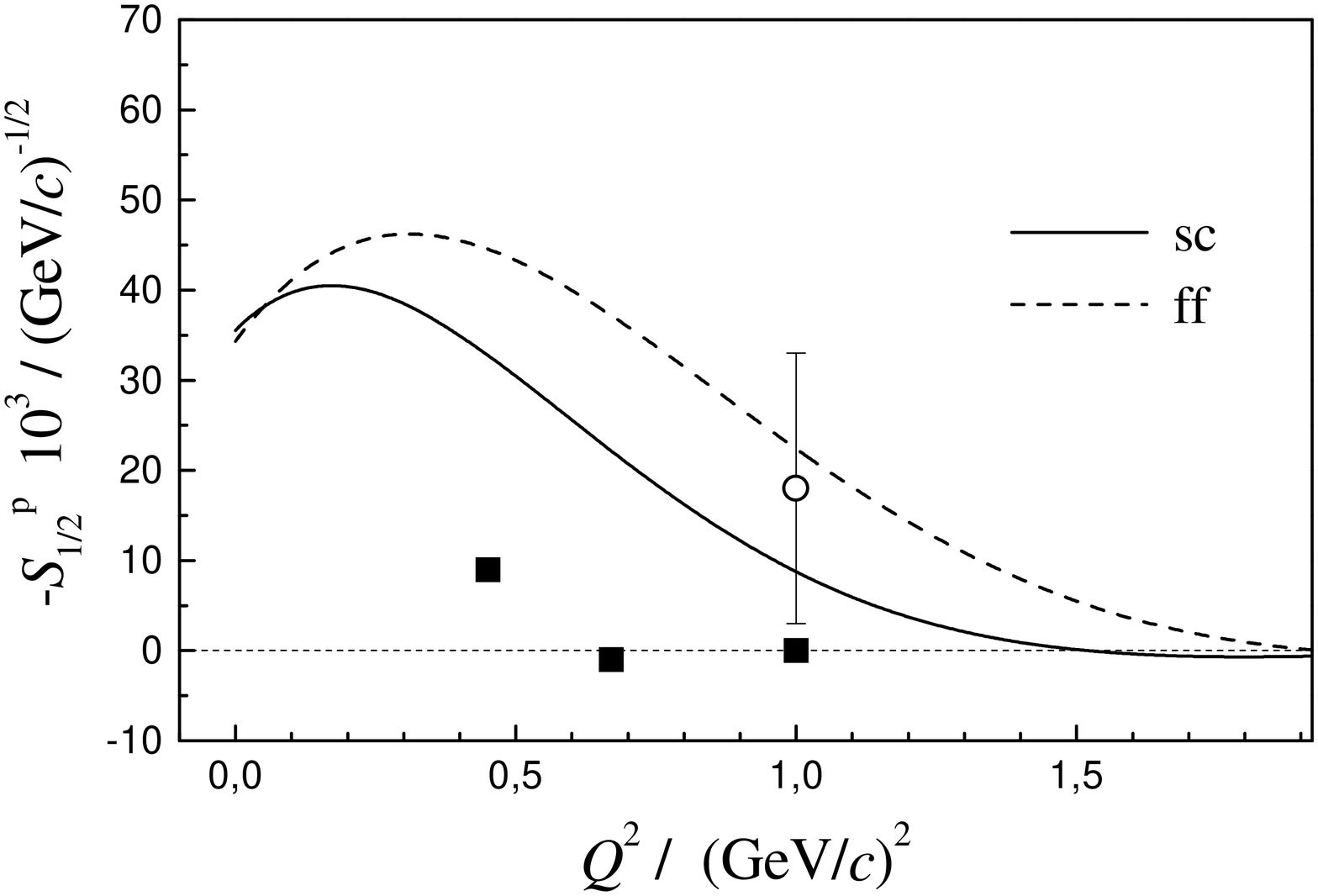,width=9.5cm} }
\caption{Nucleon Roper scalar helicity amplitude for  charged states.
See also caption of Fig.~\ref{fig:fig3}.}\label{fig:fig5}
\end{figure}

In Fig.~\ref{fig:fig5} we present 
the scalar helicity amplitudes for the proton;
the amplitude for the neutron is very close to 0 and is not shown.
No data are available which prevents 
any judgment of the quality of the approaches.

In CQM calculations~\cite{Capstick,Cardarelli,Cano}
which incorporate a consistent relativistic treatment of
quark dynamics,
the amplitudes change the sign around $Q^2\sim 0.2$--$0.5$~(GeV$/c)^2$.
The amplitudes with this opposite sign remain large at relatively high
$Q^2$, though,
as shown in \cite{Cardarelli,Cano}, the behavior at high $Q^2$
can be substantially reduced 
if either corrections beyond the simple Gaussian-like ansatz
or pionic degrees of freedom are included in the model.
Other models, in particular those including exotic (gluon) states,
do not predict this type of behavior~\cite{Li}.
The present experimental situation is unclear.
If in the future a more accurate experimental analysis confirms
the change of the sign at low $Q^2$, this would certainly be
a success of the CQM;
if, however, this is not the case, 
one should not
rule out conventional quark model explanations in favor of the
exotic states as proven by our calculation.
Our model, similarly as other chiral models~\cite{dong,tiator},  predicts
the correct sign at the photon point, while
it does not predict the change of the sign at low  $Q^2$.
Let us also note that with the inclusion of a phenomenological
three-quark interaction Cano {\em et al.} \cite{Cano} shift the change
of the sign to $Q\sim1$~(GeV$/c)^2$ beyond which, in our opinion,
predictions of low energy models become questionable anyway.

It is interesting to note that, in the self-consistent calculation,
there is a substantial contribution to the amplitudes from  
the admixures of $|{\rm N}\rangle$ in $|\tilde{\rm R}\rangle$ and of 
$|{\rm R}' \rangle$ in $|\tilde{\rm N}\rangle$ 
(see expressions (\ref{gcmss})).
Such contributions are
not present in the calculation with frozen profiles 
(since the states (\ref{usa}) are already orthogonal) but
nonetheless, both approaches yield similar results
for the amplitudes, indicating that the results are not
very sensitive to small variations of the profiles.
We should stress that we have made no attempt to fit
the electroexcitation amplitudes nor the excitation energy
of the Roper resonance but have used model parameters that were 
fixed in the ground state calculation.

We do not give 
the amplitudes for
the second excited state $N(1710)$.
As we have already 
mentioned, 
provided that the $\sigma$-meson mass is sufficiently small, 
this state 
can be dominated by a component
carrying one quantum of $\sigma$-meson vibration.
Such a picture predicts very small production amplitudes
since mostly the scalar fields are excited.
The 
presence
of $\sigma$-meson vibrations 
is consistent with the recent phase shift
analysis by Krehl {\em at al.}~\cite{Krehl} who found that
the resonant behavior in the P$_{11}$ channel can be explained
solely through the coupling to the $\sigma$-N channel,
without assuming any internal ({\em i.e.} quark) radial excitation
of the nucleon.
In our view, radial excitations of quarks are needed in order
to explain relatively large electroproduction amplitudes,
which would indicate that the $\sigma$-N channel couples to all 
nucleon $\half^+$ excitations rather than be concentrated
in the Roper resonance alone.

To conclude, though our model 
gives only a qualitative
picture of the lowest nucleon radially excited states and
their electroproduction amplitudes, it yields
some interesting features, in particular the possibility
of $\sigma$-meson vibrations,  not present in other calculations.
Its main advantage over other approaches is that all properties,
including the EM amplitudes and the resonance decay, can be
calculated from a single Lagrangian without introducing
additional assumptions; it also allows us to exactly treat 
the orthogonalization of states which is particularly important
in the description of nucleon radial excitations.
It would be instructive to check our predictions in 
other chiral models
and extend the present calculation to
include other radially excited states  such as radial
excitations of the $\Delta$.

This work was supported by FCT (POCTI/FEDER), Lisbon, and by 
The Ministry of Science and Education of Slovenia.
We thank S.~\v{S}irca and M.~Rosina for useful discussions.

\end{document}